\documentclass[english,aps,pra,reprint,noshowpacs,superscriptaddress]{revtex4-1}   
\usepackage[T1]{fontenc}	
\usepackage[latin9]{inputenc}	
\usepackage{geometry}		
\usepackage{graphicx}
\usepackage[above,below]{placeins}	
\usepackage{times}
\usepackage{hyperref}  
\hypersetup{colorlinks=true,urlcolor=blue,citecolor=blue,linkcolor=blue}   
\begin{document}

\title{What counts in laboratories: toward a practice-based identity survey}
\author{Kelsey Funkhouser}
\affiliation{Department of Physics \& Astronomy, Michigan State University, East Lansing, MI, 48824}
\author{Marcos D. Caballero}
\affiliation{Department of Physics \& Astronomy, Michigan State University, East Lansing, MI, 48824}
\affiliation{CREATE for STEM Institute, Michigan State University, East Lansing, MI, 48824}
\affiliation{Department of Physics \& Center for Computing in Science Education, University of Oslo, N-0316 Oslo, Norway}
\author{Paul W. Irving}
\affiliation{Department of Physics \& Astronomy, Michigan State University, East Lansing, MI, 48824}
\author{Vashti Sawtelle}
 \affiliation{Lyman Briggs College, Michigan State University, East Lansing, MI, 48824}
\affiliation{Department of Physics \& Astronomy, Michigan State University, East Lansing, MI, 48824}


\begin{abstract}
An essential step in the process of developing a physics identity is the opportunity to engage in authentic physics practices - an ideal place to gain these experiences is physics laboratory courses. We are designing a practice-based identity survey to be used in physics laboratory courses. A first step in determining the impact of these physics practices is understanding student's interpretations of them. In physics education research, discussions of  physics practices, are typically grounded in definitions from experts. Our students are not necessarily experts so, asking questions about what these practices mean to the students and what counts is fundamental to insure that our survey questions are being interpreted correctly. 
\end{abstract}

\maketitle

\section{Introduction}
Laboratory courses in the undergraduate physics curriculum are intended to provide students with opportunities to participate in authentic physics practices. There have been national calls to develop courses that emphasize physics practices and focus on how science is done \cite{olson2012engage,national2013next,AAPT2014}. The American Association of Physics Teachers (AAPT) Committee on Laboratories released lab guidelines, with focus areas that highlight the skills and practices necessary in doing physics \cite{AAPT2014}.

The emphasis on practices in these lab courses makes them an ideal place to investigate the impact the practices are having on our students. The opportunity to engage with authentic physics practices is an essential step in the development of a physics identity \cite{Close2013}. We are working to develop a survey to measure students' physics identity in lab courses with a focus on the effect of lab practices. Identity is a construct that is nuanced in qualitative studies, so a survey designed to probe identity must be tightly connected to a theoretical framework. In this paper we show how we build on existing survey design methods by staying close to student understanding of physics practices and add an intentional focus on the identity framework in every step in our survey design. We show how asking students to reflect on scientific practices and values  are critical steps in designing a survey on practice-based identity.
%
%
%
\section{Theoretical Background \label{sec:theory}}
\begin{figure}[htbp]
\includegraphics[clip, trim=5 200 8 150, width=0.8\linewidth]{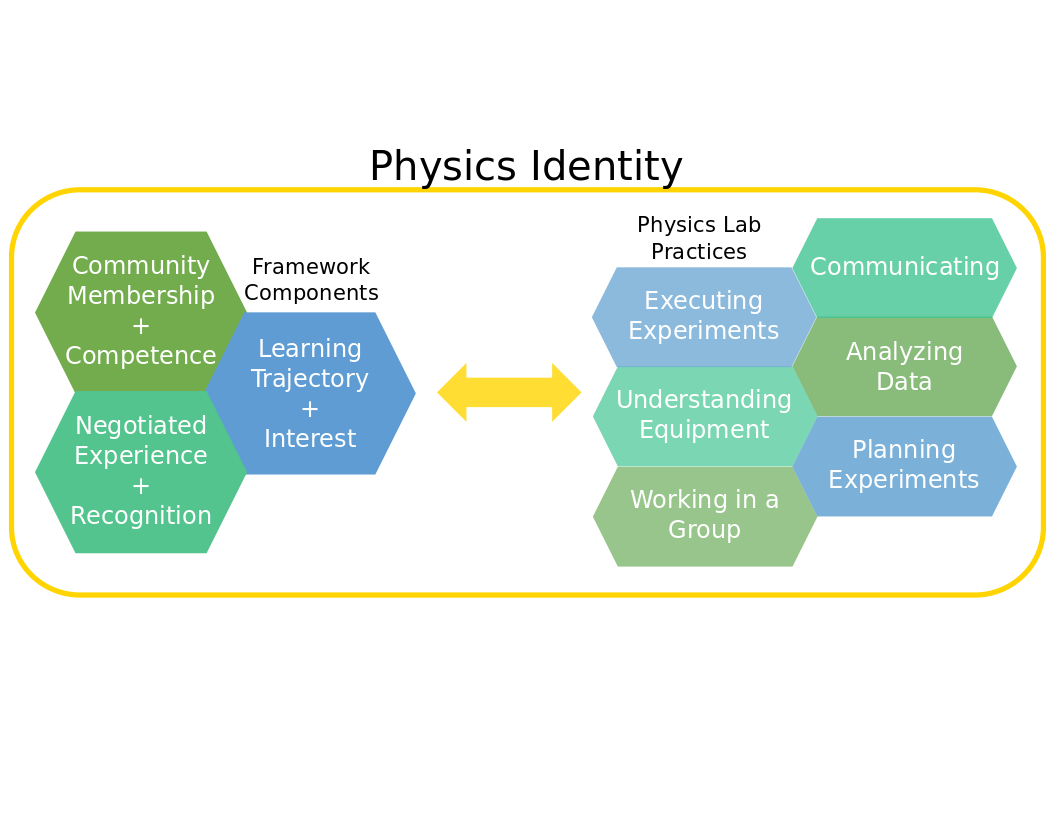}
\caption{The components of our theoretical framework, with the interaction between the physics lab practices and the components of the identity framework forming the physics identity \cite{Close2013,wenger1999communities,AAPT2014}.\label{fig1}}
\end{figure}

\textbf{Identity:} Identity is a multidimensional construct that describes how an individual sees themselves and how they situate themselves with respect to a specific community. We are adapting the identity framework presented in \cite{Close2013}, which draws from Communities of Practice (CoP) \cite{wenger1999communities} and \citet{Hazari2010} frameworks. The framework has three main components: community membership \& competence (CMC), learning trajectory \& interest (LTI), and negotiated experience \& recognition (NER) (Fig.~\ref{fig1}). Due to limits in space we will break down these components further in future publications.  For the purposes of this paper we are adapting these components to our context with the knowledge that they may not be what best describes identity in physics labs.
%
%
The context of physics labs and our goals of staying close to the theory and to the students' ideas align to pull the majority of our focus onto the specific lab practices, how students interpret them and how they identify with respect to those practices. This means examining the overlap of the experimental physics community with the other communities the students are a part of, and specifically, understanding how they interpret physics practices through the lens of those communities.

Our study of identity is further complicated by the fact that we are working to produce a survey that can be distributed on a large scale with closed response questions that combine to tell us about the participant's physics identity. This has meant that we have a carefully operationalized definition for what we count as an identity statement. We are looking for three main components of an identity statement: \textit{the student addresses the practice, applies a value statement to it, and then connects it back to one of the identity components in Fig.~\ref{fig1}.}  The definition here relies on the student's interpretation of the physics practice and a personal value statement about the practice in order to meet the first two criteria. The third component helps us to maintain our ties to the theory. We recognize that this process of operationalizing a complicated construct into a single sentence loses much of the nuance. Below, we argue that the intentional and focused process we are going through to develop question statements respects the complexity of identity as a construct.

\textbf{Physics Lab Practices:} The CoP framework on identity defines a CoP based upon the shared practices, skills, and ideas within a specific community \cite{wenger1999communities}. Physics lab practices are the skills and activities valued and utilized within the experimental physics community. We focus our initial survey design on practices from the AAPT lab guidelines and those emphasized in the physics lab courses at our institution. These initial practices were adapted based on responses in the student interviews, which are described in Sec. III A. We have thought carefully about how to represent the practices in a way that they might resonate with even if they are not intending to become physicists.

The final six broad lab practices on which we focus include: analyzing data, communicating, working in a group, understanding equipment, designing experiments, and executing experiments (Fig.~\ref{fig1}). Starting with broader definitions we have iterated on this list based on student interviews where we took in their interpretations. We acknowledge that our list of practices is limited and may not be consistent across institutions, for example our lab courses do not currently emphasize computation, while other institutions do \cite{beichner2010labs}. We hope to expand our data collection to additional institutions in order to account for this. In this paper we also do not discuss the centrality of particular practices as the CoP framework would call for, though we have structured our data collection to allow for this analysis in the future.
\section{Survey Development}
\begin{figure}
\includegraphics[clip, trim=10 10 10 10, width=0.8\linewidth]{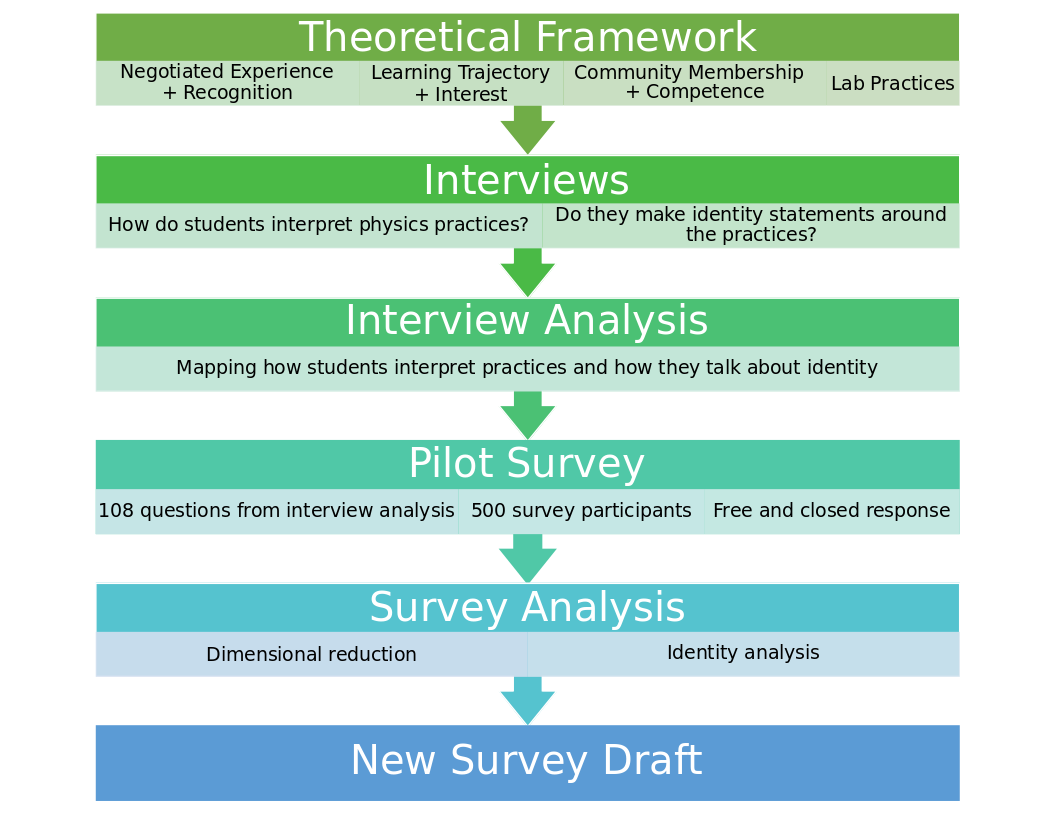}
\caption{The survey development process. The theoretical framework as described in Sec. II.\label{fig2}}
\end{figure}

We have been keeping the interpretations of the students and the components of the theoretical framework at the center of our entire survey design process (as laid out in Fig.~\ref{fig2}). In the interviews, we worked to flesh out their understandings of the lab practices that are valued in the physics community and to evaluate the ways they identify with those practices. The creation of the survey has also centered on the idea that though students may identify with physics many of them would not now, or maybe ever, identify themselves as physicists. This fact means that we need to carefully consider the intersection of students' communities with the physics community embodied in the laboratory classroom.
Therefore, in initial iterations of this survey we take into account the many communities that these students are members of: their major, the science community, sports/teams, racial/ethnicity groups etc. We acknowledge that students from life sciences are not necessarily going to see physics practices the same way that a graduating physics major might, but we assert that they both identify with physics in some way. To account for different interpretations and for future exploration of the centrality of these different practices for students we have collected data from introductory to upper division labs.
\subsection{Interviews}\label{sec:interviews}
Traditionally surveys are designed around ``expert'' ideas, but our students are not yet experts \cite{redish1998student,adams2006new}. To make sure that our questions are representative of their ideas we need to understand all the ways students are interpreting the practices. For example, one student's interpretation of ``analyzing data'' might be differ significantly from another's, which means we have to be specific about the language and interpretations we are using in the survey design. To design as survey to be applicable across a variety of students, we conducted semi-structured interviews with nine students from an introductory algebra-based physics lab for non-majors and six students in upper-division physics labs.

In the interviews we asked participants to define each of the lab practices and to describe what the practices looked like in their physics lab course (e.g., \textit{ "What does data analysis mean in your lab class?"}). Then they were asked whether or not they found that practice valuable (e.g.. \textit{"Do you find that useful?"}). This enabled us to first, explore a variety of interpretations of specific practices and second, to explore how students may (or may not) identify with that practice. 

\begin{figure}
\includegraphics[clip, trim=30 200 30 50, width=0.8\linewidth]{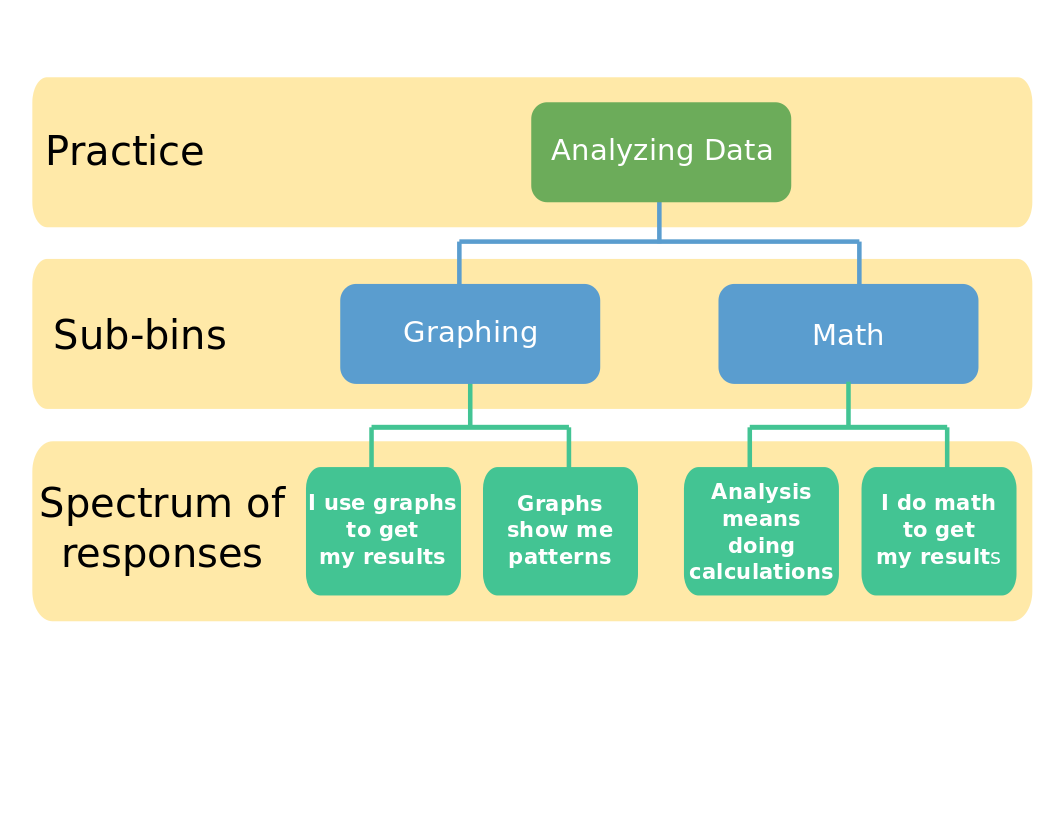}
\caption{This is an example of the interview analysis process for the practice of analyzing data.\label{fig3}}
\end{figure}

We used the interviews to map the breadth of students' interpretations of the practices, not to make claims about how specific students understand the practices. The analysis process focused on covering all interpretations and levels of centrality in the students' statements. Within each practice we found interpretations fell into specific themes, which we refer to as sub-bins (Fig.~\ref{fig3}). From those sub-bins, we identified different types of student responses with different levels of centrality -- the spectrum of responses. We then turned those responses into question stems for the pilot survey. For example,  from the interview question: \textit{What does data analysis look like in your lab?}, we looked at student responses that fell into a similar theme, graphing in this case.
\begin{quote}
As soon as you have the graph and data in front of you, with a certain mindset, [you] can interpret what it actually means. (Dean, pseudonym)
\end{quote}

\begin{quote}
So interpreting the results, in this class, we use graphs and stuff to do that. (Fern)
\end{quote}

Looking across responses that fell into similar categories, we wrote pilot question statements that would represent student interpretations of the practices that they could endorse in closed response questions. For example, the above student quotes combine into the question stem for the pilot survey: \textit{In my lab, using graphs to interpret results is important.} An overview of the practice space and the pilot survey questions produced can be seen in Table.~\ref{tab1}.
\begin{table}[tbp]
\caption{Overview of Practice Space.\label{tab1}}
\begin{ruledtabular}
\begin{tabular}{lcc}
 \textbf{Practice}&\textbf{Sub-bins}&\textbf{Pilot Survey Questions}\\ 
 \hline
Analyzing Data  &7&19 \\
Communicating & 3&18\\
Working in a Group &9&25\\
Understanding Equipment &7&25 \\
Planning Experiments &3&14 \\
Executing Experiments &2&7 \\
 \hline
Total & 31 &108
\end{tabular}
\end{ruledtabular}
\end{table}
%
%

In the analysis we also evaluated the ways students described how they identified with specific practices. This lead us to our definition of an identity statement as discussed in Sec.~\ref{sec:theory}. To highlight the steps we went through to form our definition we compare two statements where students identify with lab practices.
\begin{quote} Blake(discussing uncertainty): ...I do understand its importance, and see it as a necessary thing. It's just, I personally will never really need to take uncertainty as seriously as it is taken in these upper level lab classes and...I don't look at it the same way as someone who's thinking `I want to do experimental physics for the rest of my life'. \end{quote}

We see that Blake refers to a specific practice, uncertainty in analyzing data (which he defines earlier in the interview). He makes a \textit{personal} value statement about it: \textit{I do understand its importance, and see it as a necessary thing}. He then connects it back to the identity framework: \textit{I personally will never really need to take uncertainty as seriously.} We see this statement fitting into the LTI component of the identity framework (Fig.~\ref{fig1}) because Blake is projecting forward to his future career (as a high school physics or math teacher), where he does not think he will need to use uncertainty. These components of Blake's statement address our main goals in the survey development process, covering his interpretation of the practice, the way he does (or does not) identify with it, and then connecting it back to our theoretical framework. We contrast this example with one that does not make the connection back to the theory.
\begin{quote} Beth (describing something she found exciting): It was good to, plug it all into the calculator and [see] this is a reasonable answer. So, getting something that...could be right and doing that without a procedure is more self-satisfying because...you relied more on yourself.  \end{quote}

Beth, too, refers to a practice, broadly analyzing data, and she applies a value statement to it: \textit{more self-satisfying because...you relied more on yourself.} But, she does not make the value statement in a way that easily connects to the framework to mark it as an identity statement. If she had talked about how important self-reliance is either for her future (LTI) or as an important part of a community she is a part of (CMC), then we could classify it as an identity statement. This comparison between statements from Blake and Beth exemplifies how we came to include the third requirement to our identity statement definition, the students' statement connects back to an identity component in (Fig.~\ref{fig1}). 

The analysis of the interview data supported covering the breadth of students' interpretations of the six lab practices (Table \ref{tab1}). Working from this analysis, we built questions for the pilot survey that explicitly asked students to reflect on the practice and the value statement, but left open how participants' statements connect the practice back to the identity framework. 
%
%
%
\subsection{Pilot Survey}
The purpose of the pilot survey was not statistical validation that is traditionally part of a survey development process, we were not yet at a point where such validation would be necessary or useful. The overarching goal instead was to cover the ways that students identify with the physics practices, similarly to how the interviews were used to determine how students were interpreting the practices. This connects well with our intentions in the design process, we want to be certain of our students' ideas and interpretations, and we need more information  to create questions to probe identity reliably. In addition to the identity information we are using the pilot survey to confirm full coverage of the practice space. 
 
We worked to elicit identity statements that met our criteria and we primed participants to think about specific communities throughout the survey. We built the first two components of the criteria directly into the prompt, addressing a practice and applying a value statement to it. We then broke each question statement into four pieces. The first two were Likert-scale closed response items, which gave the participants space to endorse (or not) the value statements. The other two were free response questions where participants were asked to explain their reasoning,  the aim being that they would make that final connection to the identity framework. 

\begin{figure}
\includegraphics[clip, trim=50 195 50 150, width=0.8\linewidth]{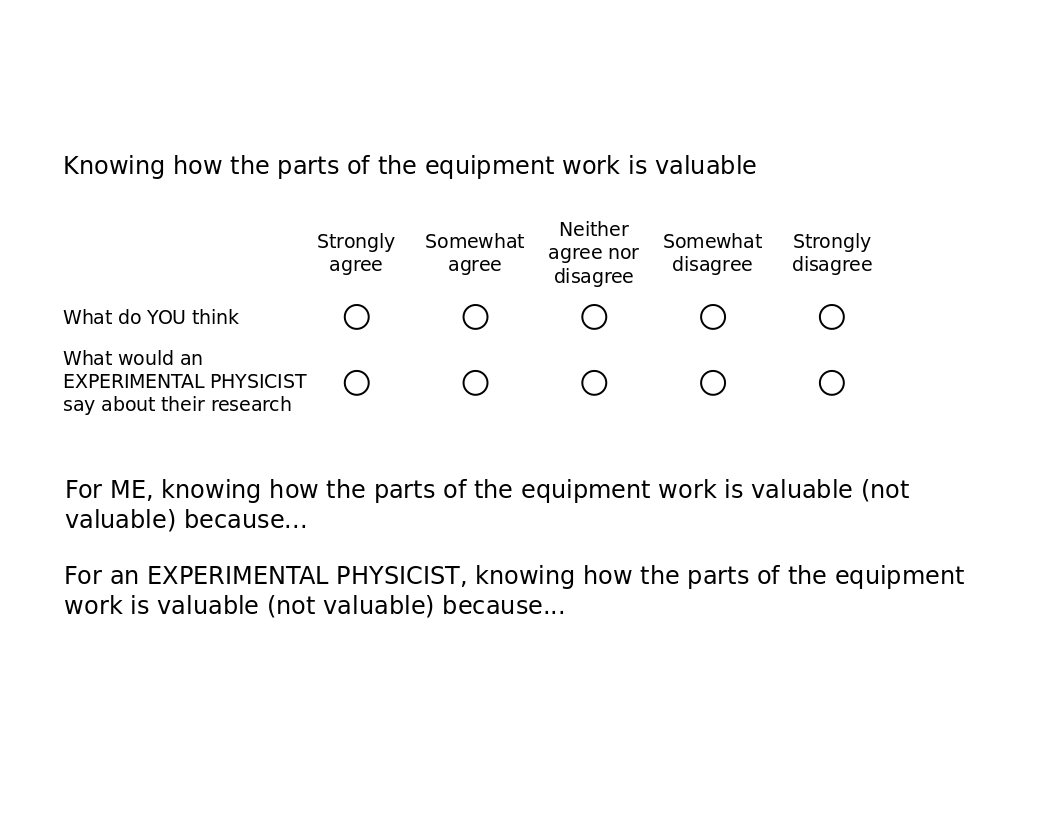}
\caption{An example question from the pilot survey.\label{fig4}}
\end{figure}

In each question group we asked the participants for both their own perspective and what they thought an experimental physicist might say (see Fig.~\ref{fig4} for an example), following the strategy from the Colorado Learning About Science Survey in Experimental Physics (ECLASS) \cite{Zwickl2014}. We did this as a way to prime them to think about specific communities when answering questions, in addition to asking about the communities they fell into at the beginning and end of the survey. We also found that in preliminary trials participants supplied the more ``expert'' response before the more personal one, so this provided them space to do both in a transparent way.\\
To ensure that our survey development process remained aligned with student ideas we needed to retain the level of variation in their interpretations of the practices.  We structured the pilot survey to target this alignment by first, maintaining our large sample of question statements from the interview analysis. We also needed large numbers in responses to justify removing redundant questions and to confirm that our interviews were representative of a much larger sample. This was accomplished by distributing the survey to the students in the introductory algebra-based lab for non-majors at the end of the spring semester, roughly 650 students. With nearly 500 responses to the survey we received between 40-60 individual responses to each of the 108 questions.

\section{Discussion}

This paper has described the process we have taken in developing a practice based identity survey, with our ideas about identity and practices grounded in the literature \cite{Close2013,wenger1999communities,AAPT2014}.  We described our process of using interviews to map the breadth of students' interpretations of the practices. From that, we produced 108 questions for an atypical pilot survey, where we work to validate the interpretations of practices and elicit identity statements. 

We argue that in order to produce a robust survey that measures our intended construct it is essential to make sure each component of the survey development process is focusing on the students, derived from the information we get from them and is closely tied to the theoretical framework on identity. In doing so we intend to produce a survey that will tell us about our students' physics identities and the practices we emphasize in our lab courses. We also aim to progress in our understanding of physics identity, especially for students who would not necessarily call themselves physicists or those who do not intend to go into a physics related field at all. We posit that a process like ours is necessary to ensure a survey speaks to and for the community we are studying, students in physics lab courses in our case. Especially, when studying something as nuanced as identity, we need to be intentional every step of the way to understand our students, to ensure that they understand us, and to retain the theory as the foundation.
\acknowledgments{We thank the Howard Hughes Medical Institute for financial support as well as Michigan State University's Department of Physics \& Astronomy for their continued support of transformed introductory labs. We thank undergraduate researcher Carissa Meyers, for her invaluable efforts in analysis of the pilot survey.}

\bibliographystyle{apsrev}  	
\bibliography{sample.bib}  	

\end{document}